\documentclass[aps,twocolumn,amsmath,amssymb]{revtex4}
\usepackage{graphicx,color}
\pdfoutput=1
\usepackage{dcolumn}
\usepackage{hyperref}

\let\emptyset\varnothing

\begin{document}
\title{Topological vertex formalism with O5-plane}

\author{Sung-Soo Kim}\email{sungsoo.kim@uestc.edu.cn}
\author{Futoshi Yagi}\email{fyagi@physics.technion.ac.il}
\affiliation{School of Physical Electronics, University of Electronic Science and Technology of China, North Jianshe Road, Chengdu 611731, China}
\affiliation{Department of Physics, Technion - Israel Institute of Technology, Haifa 32000, Israel}

\begin{abstract}
We propose a new topological vertex formalism for a type IIB $(p,q)$ 5-brane web with an O5-plane. We apply our proposal to five-dimensional $\mathcal{N}=1$ Sp(1) gauge theory with $N_f=0,1,8$ flavors to compute the topological string partition functions and check the agreement with the known results. Especially for the $N_f=8$ case, which corresponds to E-string theory on a circle, we obtain a new, yet simple, expression of the partition function with a two Young diagram sum. 
\end{abstract}

\maketitle


\section{Introduction}\label{sec:intro}
The topological string partition function has played an important role in finding the Bogomol'nyi–Prasad–Sommerfield (BPS) spectrum (e.g., Nekrasov partition function) of the supersymmetric gauge theories with eight supercharges. Given a generic toric Calabi-Yau geometry, the partition function can be systematically computed based on topological vertex formalism \cite{Aganagic:2003db, Iqbal:2007ii, Awata:2008ed}.
 The correspondence \cite{Leung:1997tw} between the toric diagram and the $(p,q)$ 5-brane web diagram allows one to implement the formalism  on a $(p,q)$ 5-brane web. 
Recently, a generalized version of the 5-brane web diagram, corresponding to nontoric Calabi-Yau geometry, was introduced in \cite{Benini:2009gi}, 
and it has been proposed \cite{Hayashi:2013qwa, Hayashi:2014wfa, Hayashi:2015xla} that topological vertex formalism is applicable even for such nontoric cases simply by tuning K\"ahler parameters in a proper way 
\cite{Dimofte:2010tz, Taki:2010bj, Aganagic:2011sg, Aganagic:2012hs}.

\begin{figure}
\centering
\includegraphics[width=7.5cm]{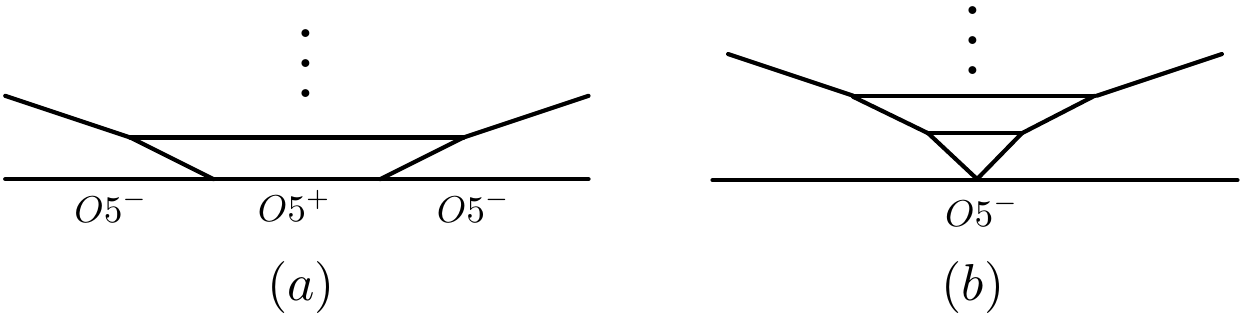}
\caption{(a) 5-brane web diagram for five-dimensional $\mathcal{N}=1$ Sp($N$) gauge theory. (b) 5-brane web diagram for Sp($N$) { with $\theta=0$ (for odd $N$) or with $\theta=\pi$ (for even $N$)} after the general flop transition, where Sp($N$) theory description is not valid. Instead it describes five-dimensional $\mathcal{N}=1$ SU($N+1$) gauge theory with $\kappa=N+3$. 
}
\label{Fig:web-Sp}
\end{figure}

It is then natural to ask whether topological vertex formalism can be also applicable to the $(p,q)$ 5-brane web diagram whose corresponding Calabi-Yau geometry is not necessarily clear
\footnote{At least, the opposite case, where Calabi-Yau geometry is known but 5-brane configuration is not known, is discussed in~\cite{Hayashi:2017jze}. 
There is also vertex-like approach~\cite{Diaconescu:2005ik,Diaconescu:2005tr,Diaconescu:2005mv}.
}.  
In this paper, we discuss that it is possible at least for some cases. 
In particular, we consider a type IIB $(p,q)$ 5-brane web with an O5-plane describing five-dimensional $\mathcal{N}=1$ Sp($N$) gauge theory, as depicted in Fig. \ref{Fig:web-Sp}(a) \cite{Zafrir:2015ftn, Hayashi:2017btw}. The five-dimensional pure Sp($N$) theory has a $\mathbb{Z}_2$ valued discrete theta angle, either $\theta=0$ or $\pi$ \cite{Morrison:1996xf, Douglas:1996xp}. In Fig. \ref{Fig:web-Sp}(a), however, the difference in the theta angles does not look manifest \cite{Zafrir:2015ftn}. It becomes much more distinct when one uses a generalized flop transition developed in \cite{Hayashi:2015vhy, Hayashi:2017btw}. For instance, Fig. \ref{Fig:web-Sp}(b) is the web diagram for the discrete theta angle $\theta=0$ (for odd $N$) after the flop transition is performed on Fig. \ref{Fig:web-Sp}(a), 
where  $(1,1)$ and $(1,-1)$ 5-branes are attached to an O5-plane in a specific way 
\footnote{For the discrete theta angle $\theta=\pi$, one obtains a different brane configuration after the flop transition. For details, see \cite{Hayashi:2017btw}.}.
In this paper, we propose new rules for such configuration in addition to the conventional topological vertex formalism, and present a new method that enables one to compute the Nekrasov partition function for such gauge theory constructed with $(p,q)$ 5-branes with an O5-plane
\footnote{There were other topological vertex formulations in the presence of orientifold planes, for instance, the real topological string \cite{Walcher:2007qp, Krefl:2009md, Krefl:2009mw}. 
Note, however, that these formalisms are different from ours, as these orientifolds are codimension two from the gauge theory point of view \cite{Piazzalunga:2014waa, Hayashi:2015uka}, while our case rather corresponds to codimension zero, and thus the gauge theories are different.}.

 It is worth noting that the web diagram in Fig.
\ref{Fig:web-Sp}(b) corresponds to the parameter region where the gauge coupling square is negative, which is denoted in \cite{Aharony:1997bh} as ``past infinite coupling.''  
In this region, the description as five-dimensional Sp($N$) gauge theory breaks down. Instead, better description is given by five-dimensional SU($N+1$) gauge theory with Chern-Simons level $\kappa = N+3$. This duality is first proposed in \cite{Gaiotto:2015una} with $N_f$ flavors and $\kappa = N+3-N_f/2$, and further elaborated in \cite{Hayashi:2015zka, Hayashi:2015vhy, Hayashi:2016abm}.  
The map between the parameters of two gauge theories are proposed in \cite{Gaiotto:2015una, Hayashi:2016abm} and it is checked that the Nekrasov partition functions for the dual theories are identical under this map up to analytic continuation. Due to the nontrivial Chern-Simons level, the Nekrasov partition function for this five-dimensional SU($N+1$) gauge theory has been less understood, although there are several partial results including~\cite{Gaiotto:2015una, Hayashi:2016abm}. Throughout the paper, we demonstrate our topological vertex formalism with an O5-plane would provide an effective way for computing Nekrasov partition function of this class. To this end, we focus on the simplest case, $N=1$, as an instructive example of our proposal. Since Sp(1)$\,=\,$SU(2), the map gives symmetry of the partition function, which in fact corresponds to a Weyl transformation of the enhanced $E_{N_f+1}$ global symmetry~\cite{Hayashi:2016abm}. 

The organization of the paper is as follows: In Sec.~\ref{sec:formalism}, we propose a new rule of topological vertex formalism for the intersection between 5-branes and an O5-plane appearing in Fig.~\ref{Fig:web-Sp}(b) as well as still another type of intersection which also naturally appears through generalized flop transition. In Sec.~\ref{sec:computation}, we compute (unrefined) Nekrasov partition function for SU(2) $N_f=0,1$ and $N_f=8$ flavors based on the new formalism introduced in Sec.~\ref{sec:formalism}, and compare the obtained result with the known one. In particular, the case with $N_f=8$ flavors corresponds to the E-string partition function. We then conclude with summary and future directions.

\section{Formalism}\label{sec:formalism} 
In this section, we propose new rules for computing topological string partition function that involves a brane configuration  with an O5-plane, in addition to the conventional (unrefined) topological vertex formalism. For notations, we follow \cite{Bao:2013pwa} mostly.

According to the topological vertex formalism \cite{Aganagic:2003db}, the topological string partition function can be computed systematically 
based on the $(p,q)$ 5-brane web diagram as follows:
First, we assign different Young diagrams $\lambda, \mu, \nu, \cdots$ to different edges in the web diagram. Then, we introduce the edge factor
$(-Q)^{|\lambda|}\,\, f_\lambda{}^{\mathfrak{n}}$ to each edge,
where $\lambda$ is the Young diagram assigned to the considered edge,
and the vertex factor $C_{\lambda\mu\nu}$ is assigned to each vertex,
where $\lambda, \mu, \nu$ are the Young diagrams assigned to the three edges sharing the vertex we consider.
The Young diagrams in the edge factor are ordered in a clockwise way
\footnote{
It is known \cite{Aganagic:2003db, Okounkov:2003sp} that the unrefined vertex factors are invariant under cyclic permutation 
$$
C_{\lambda\mu\nu}=C_{\mu\nu\lambda}=C_{\nu\lambda\mu}.	
$$
}.
Here, $Q$ is the exponentiated length of the 5-branes,
which corresponds to the K\"ahler parameter of the corresponding toric Calabi-Yau geometry.
The framing factor $f_\lambda$ 
\footnote{
The explicit form of the framing factor associated with an edge is given by 
$$
f_\lambda(g)
=(-1)^{|\lambda |} 
g^{\frac{||\lambda^t||^2-||\lambda||^2}{2}} 
~\big(=f^{-1}_{\lambda^t}(g)~\big),
$$ 
with $|\lambda| = \sum^{\ell(\lambda)}_{i=1}\lambda_i$ for Young diagram  $\lambda=(\lambda_1, \lambda_2,\cdots, \lambda_{\ell({\lambda})})$ assigned to the edge, and 
$||\lambda||^2 = \sum^{\ell(\lambda)}_{i=1}\lambda_i^2$.}
is a specific function of $g= e^{-\beta\epsilon}$, with $\epsilon$ being the self-dual $\Omega$-deformation parameter $\epsilon \equiv \epsilon_1=-\epsilon_2$.
The power $\mathfrak{n}$ is fixed by the $(p,q)$ charge of the adjacent 5-branes.
See, e.g.~\cite{Bao:2013pwa} for detailed explicit expressions.
Finally, $C_{\lambda\mu\nu}$ is the (unrefined) topological vertex defined in \cite{Aganagic:2003db}, which is a specific function of $g$ written in terms of the skew Schur functions
\footnote{
The explicit expression is given by
$$
C_{\lambda\mu\nu} = g^{\frac{-||\mu^t||^2+||\mu||^2 +||\nu||^2}{2}}\,
\widetilde{Z}_\nu(g)\,\sum_\eta 
s_{\lambda^t/\eta}(g^{-\rho-\nu})\,s_{\mu/\eta}(g^{-\rho-\nu^t}),
$$
with 
$$
\widetilde{Z}_\nu(g) 
=\widetilde{Z}_{\nu^t}(g) =
\prod_{i=1}^{{\ell}(\nu)}\prod_{j=1}^{\nu_i}\frac1{1-g^{\nu_i+\nu^t_j-i-j+1}} ,
$$
and $s_{\lambda/\eta}(x)$ is a skew-Schur function.
}.
Then the topological string partition function can be computed by
multiplying these factors and summing over all possible Young diagrams as
\begin{align}\label{eq:top}
Z = \sum_{\lambda, \mu,\nu \cdots}\,\prod\, ({\rm Edge ~factor}) \cdot \prod \,({\rm Vertex~ factor}).
\end{align}

\begin{figure}
\centering
\includegraphics[width=8cm]{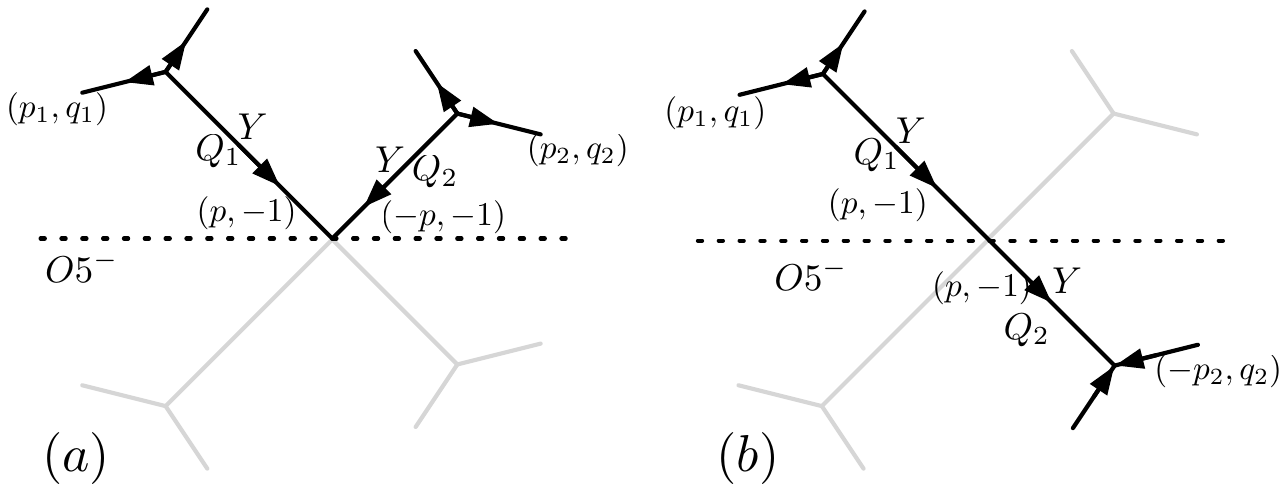}
\caption{(a) New rule for topological vertex formalism including an O5-plane. (b) Interpretation of the rule in terms of the reflected image by an O5-plane.}
\label{Fig:new1}
\end{figure}
When we have an O5-plane, we need new rules for the part where 5-branes are intersected with an O5-plane. 
Let us consider a configuration where the $(p,-1)$ 5-brane and $(-p,-1)$ 5-brane intersect on the O5$^-$-plane 
as depicted in Fig. \ref{Fig:new1}, 
where $p$ is either $p=0$ or $p=1$. The configurations for both $p=0$ and $p=1$ naturally appear as a consequence of the generalized flop transition discussed in \cite{Hayashi:2017btw}. Note that the case $p=0$ corresponds to the two coincident NS5-branes attached to an O5-plane.
The K\"ahler parameters associated with 
the $(p,-1)$ and $(-p,-1)$ 5-branes are $Q_1$ and $Q_2$, respectively.

For a $(p,q)$ 5-brane web with an O5-plane like Fig. \ref{Fig:new1}, we now introduce the following new rule for the topological vertex computations:

\begin{itemize}
	\item Assign identical
         Young diagram $Y$ to both the $(p,-1)$ 5-brane and the $(-p,-1)$ 5-brane as in Fig. \ref{Fig:new1}(a).
         
	\item Introduce the new type of edge factor, 
	\begin{align}\label{eq:O5assign}
         ( + Q_1 Q_2)^{|Y|} f_{Y}^{\mathfrak{n}}(g),
        \end{align}
        for the configuration including the edges corresponding to $(p,-1)$ and $(-p,-1)$ 5-branes,
        where
        \begin{align}\label{eq:framingO}
        \mathfrak{n} = (p_2, - q_2) \wedge (p_1,q_1) + 1 = p_1 q_2 + p_2 q_1 + 1.
        \end{align}
        \end{itemize}
Equipped with these two new rules, we claim that the topological string partition function can be computed in the same way as in \eqref{eq:top} even if the 5-brane web diagram includes O5-planes. 

These new rules would be more intuitive when we see the brane configuration from the point of view of the covering space which includes the reflected images due to an O5-plane. Namely, the configuration in Fig. \ref{Fig:new1}(a) can be naturally connect to the reflected 5-branes as shown in Fig. \ref{Fig:new1}(b). The resulting configuration then look a single edge, and it is therefore natural to assign the identical Young diagram $Y$. 
The new edge factor in \eqref{eq:O5assign} is also analogous to the conventional edge factor.

We note, however, that the direction of arrows on the edges are all inverted compared with naive expectation.
Also the $+$ sign appearing in \eqref{eq:O5assign} is different from the one appearing in the conventional edge factor. Furthermore, $+1$ is added to the power of the framing factor compared with the naive expectation from Fig. \ref{Fig:new1}(b). 

All these subtle differences appear in the following reason:
If we compare the contribution coming from the sub-diagram reflected due to an O5-plane to the original one, 
we find that the order of Young diagrams in the topological vertices should be reverse since the clockwise ordering is converted to a counter-clockwise ordering when we reflect the diagram along the O5-plane.
From the identity, 
\begin{align}\label{eq:Ct}
C_{\lambda \mu  \nu} 
= (-1)^{| \lambda | + | \mu | + | \nu |} 
f_{\lambda}^{-1}(g) f_{\mu}^{-1}  (g) f_{\nu}^{-1}  (g)
C_{\mu^t \lambda^t \nu^t},
\end{align}
we find that the reversal of the order of the Young diagram is translated into the transposition of the Young diagram, which is equivalent to changing the direction of the arrow.
Also, the prefactors in \eqref{eq:Ct} account for the $+$ sign in \eqref{eq:O5assign} as well as $+1$ in \eqref{eq:framingO}.
This reflection technique is useful for practical computations.

\section{Computation}\label{sec:computation}
In this section, we apply our proposal for the topological vertex method for a brane configuration with an O5-plane to a few specific well known examples: five-dimensional Sp(1) theory with $N_f = 0,1$ and $N_f=8$ flavors.
More specifically, we compute the BPS partition function of the five-dimensional Sp(1) theory and test our method by comparing our obtained result with the well-known SU(2) theory result. 

\subsection{\texorpdfstring{$N_f=0$}{Nf=0} case}\label{sec:Nf0}
\begin{figure}
\centering
\includegraphics[width=8.5cm]{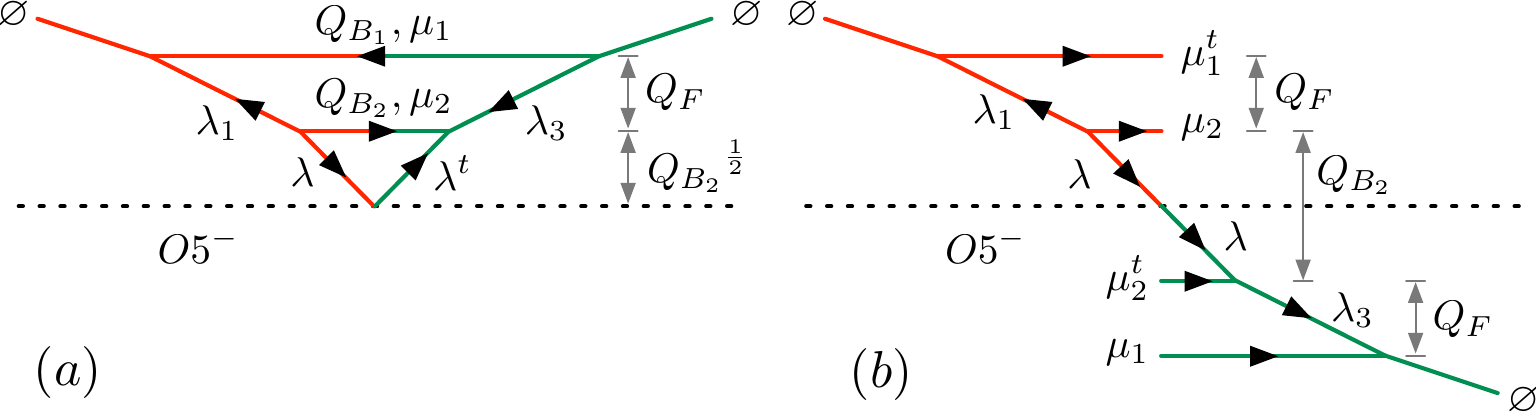}
\caption{(a) $N_f=0$ case where physical parameters are expressed in accordance with the SU(2) parametrization: $Q_F= A^2$ accounts the Coulomb branch modulus, while $Q_{B_2}Q_F=\mathfrak q$ accounts for the instanton factor, and they satisfy $Q_{B_1}= Q_{B_2}Q_{F}^4$. (b) A partial brane configuration connected to the reflected image by an O5-plane.}
\label{Fig:O5Nf0}
\end{figure}
Consider five-dimensional pure Sp(1) theory with the discrete theta angle $\theta=0$ (mod $2\pi$) whose brane configuration is given in Fig. \ref{Fig:O5Nf0} \cite{Hayashi:2017btw}.
Following the computation procedure in Sec. \ref{sec:formalism} based on Fig. \ref{Fig:O5Nf0}(a), 
we obtain the partition function as
\begin{align}\label{eq:O5_E1_top}
Z_{0}^{\rm O5} 
= &\sum_{\mu_1, \mu_2} 
(-Q_{B_1})^{|\mu_1|} f_{\mu_1}^{-5} 
(-Q_{B_2})^{|\mu_2|} f_{\mu_2}^{3} 
\cr
& 
\times \sum_{\lambda}
(Q_{B_2})^{|\lambda|} f_{\lambda}
Z^{\text{red}}_{\mu_1 \mu_2 \lambda}
Z^{\text{green}}_{\mu_1 \mu_2 \lambda},
\end{align}
where we glued the red strip and the green strip in Fig. \ref{Fig:O5Nf0}, defined as
\begin{align}
&Z^{\text{red}}_{\mu_1 \mu_2 \lambda} = \sum_{\lambda_1} (-Q_F)^{|\lambda_1|} f_{\lambda_1} C_{\lambda_1^t \emptyset \mu_1^t} C_{\lambda \lambda_1 \mu_2},
\label{eq:red}
\\
& Z^{\text{green}}_{\mu_1 \mu_2 \lambda} = \sum_{\lambda_3} (-Q_F)^{|\lambda_3|} f_{\lambda_3} C_{\emptyset \lambda_3 \mu_1}C_{\lambda_3^t \lambda \mu_2^t}.
\label{eq:green}
\end{align}
Instead of computing \eqref{eq:O5_E1_top} directly, we implement the reflection technique introduced in the previous section, which is more convenient and systematic. First we use the identity \eqref{eq:Ct} to re-express the red and green strips as a single strip as in Fig. \ref{Fig:O5Nf0}(b), which involves the summation over $\lambda$ in \eqref{eq:O5_E1_top}. We denote the resultant strip by  $Z^{\text{strip}}_{\mu_1 \mu_2}$. The full partition function is then  written as
\begin{align}\label{eq:Z0:strip}
Z_{0}^{\rm O5} = \sum_{\mu_1, \mu_2} 
(Q_{B_1})^{|\mu_1|} f_{\mu_1}^{-6} 
(Q_{B_2})^{|\mu_2|} f_{\mu_2}^{4} 
Z^{\text{strip}}_{\mu_1 \mu_2}.
\end{align}
Notice that $Z^{\text{strip}}_{\mu_1 \mu_2}$ is nothing but a conventional strip diagram contribution which is already discussed in \cite{Iqbal:2004ne}, and it is straightforward to compute
\begin{align}\label{eq:O5nf0}
Z^{\text{strip}}_{\mu_1 \mu_2} 
& = 
 \tilde{Z}_{\mu_1}^2  \tilde{Z}_{\mu_2}^2 g^{||\mu_1^t||^2+||\mu_2||^2}
\\
&\times\! \frac{ R_{\mu_2 \mu_2} (Q_{B_2})  R_{\mu_1^t \mu_2} (Q_F Q_{B_2})^2 R_{\mu_1^t \mu_1^t} (Q_F^2 Q_{B_2})}%
{ R_{\mu_2^t \mu_1^t} (Q_F)^2},\nonumber
\end{align}
where we defined
\footnote{For practical calculation, one can use the following expression
$$	R_{\lambda \mu}(Q)
	=M(Q)^{-1} \,N_{\lambda^t\mu}(Q),
$$	
where
$
M(Q) =\displaystyle\prod_{(i,j)=1}^{\infty} (1-Q\,g^{i+j-1})^{-1},
$
$$
N_{\lambda\mu}(Q) =\prod_{(i,j)\in \lambda}(1-Q\,g^{\lambda_i+\mu^t_j-i-j+1})\prod_{(i,j)\in \mu}(1-Q\,g^{-\lambda^t_j-\mu_i+i+j-1}).
$$
} 
\begin{align}
\widetilde{Z}_\nu(g)  &=
\prod_{(i,j) \in \nu} \frac{1}{1-g^{\nu_i+\nu^t_j-i-j+1}} ,
\\
R_{\lambda \mu}(Q)
&= \prod^{\infty}_{i,j=1}\left(1- Q\, g^{i+j-\mu_i-\lambda_j-1}\right).
\end{align}
Here, the K\"ahler parameters in Fig. \ref{Fig:O5Nf0} are related to
the Coulomb modulus $A=e^{-\beta a}$ and the instanton factor $\mathfrak{q}$ as follows,
\begin{align}
Q_F = A^2, \qquad Q_{B_1} = \mathfrak{q}\, A^{6}, \qquad  Q_{B_2} = \mathfrak{q} \,A^{-2}.
\end{align}
We then expand the partition function \eqref{eq:Z0:strip} in terms of $\mathfrak{q}$ to compare it with the known results \cite{Bao:2013pwa, Mitev:2014jza}. We checked their agreement up to 10 instanton orders. 

\subsection{\texorpdfstring{$N_f=1$}{Nf=1} case}
Adding a flavor D5-brane to the brane configuration is straightforward. 
\begin{figure}
\centering
\includegraphics[width=7cm]{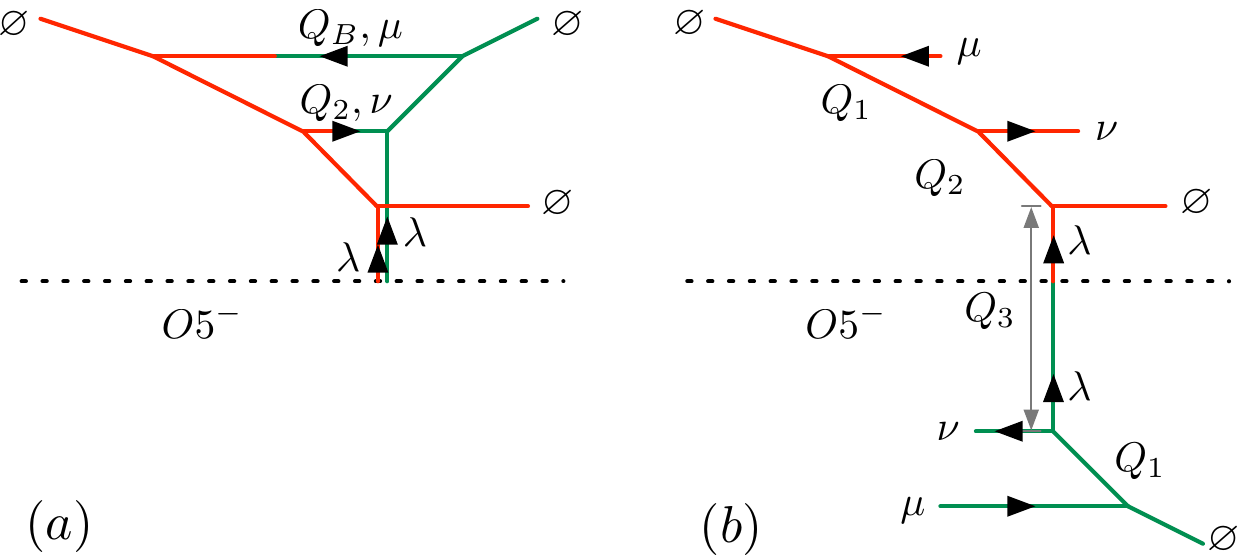}
\caption{(a) A $N_f=1$ configuration with an O5$^-$-plane. (b) A brane configuration connected by a reflected image by an O5-plane.
}
\label{Fig:O5Nf1}
\end{figure}
As a representing example, we discuss the $N_f=1$ case based on Fig. \ref{Fig:O5Nf1}, which includes the configuration in Fig. \ref{Fig:new1} with $p=0$. We note that
depending on the region of the mass parameter, one can also use a diagram with the configuration in Fig. \ref{Fig:new1} with $p=1$. 
Although the cases of $p=0$ and $p=1$ may look different at first sight, 
one can easily check that either case gives the same strip
when we separate and glue the half of the diagram to obtain Fig. \ref{Fig:O5Nf1}(b) \cite{Hayashi:2017btw}.
The only difference is either the flavor brane is placed above or below the position of the O5-plane,
which does not change the computation at all.

Repeating the procedure explained in the previous subsection, 
one readily obtains the partition function for the  $N_f=1$ case,
\begin{align}\label{eq:O5nf1}
Z_{1}^{\rm O5} & = \sum_{\mu, \nu}
(-Q_B)^{|\mu|} (-Q_2)^{|\nu|}
\\
&\times 
g^{\frac{5}{2}||\mu||^2- \frac{3}{2} ||\mu^t||^2 - \frac{1}{2} ||\nu||^2 + \frac{3}{2}||\nu^t||^2}\widetilde{Z}_{\mu}^2\widetilde{Z}_{\nu}^2 
\cr
& \times
R_{\emptyset\nu} (Q_3)  
R_{\nu\nu} (Q_{23} )
R_{\emptyset \mu^t}(Q_{13}) 
R_{\mu^t\nu}(Q_{123})^2
R_{\mu^t\mu^t} (Q_{1123})
\cr
& \times
R_{\mu^t \nu^t}(Q_1)^{-2}\,
R_{\nu\emptyset}(Q_2)^{-1}\,
R_{\mu^t\emptyset}(Q_{12})^{-1},\nonumber
\end{align}
where we used a shorthand nation for K\"ahler parameters, $Q_{ijk\cdots}=Q_iQ_jQ_kQ_{\cdots}$.
The relation between the K\"ahler parameters and  the gauge theory parameters is given as follows,
\begin{align}
	Q_1=A^2,~ 	
	Q_2=\frac{\mathfrak{q}}{A M^{\frac12}},  ~ 	
	Q_3=\frac{M}{A},~ 
	Q_B= \frac{\mathfrak{q}A^5}{M^{\frac12}}, 
\end{align}
where $A$ is the Coulomb modulus, $M=e^{-\beta m}$ is the mass parameter, and $\mathfrak{q}$ is the instanton factor.
Again, by expanding \eqref{eq:O5nf1} in terms of $\mathfrak{q}$, we checked our method. Our partition function \eqref{eq:O5nf1} agrees with the known result \cite{Bao:2013pwa, Mitev:2014jza} up to 10 instantons, 
where we used the flop transitions in the perturbative part. In a similar fashion, 
the partition functions for the $N_f=2,\cdots,7$ cases can be straightforwardly computed. 
\subsection{\texorpdfstring{$N_f=8$}{Nf=8} case: E-string theory}
We now consider the $N_f=8$ case which would serve as a nontrivial test for our method. The five-dimensional SU(2) theory with $N_f=8$ flavors is special in the sense that its UV fixed point exists in six dimensions. It is in fact six-dimensional E-string theory compactified on a circle whose partition function or elliptic genus was recently computed in \cite{Hwang:2014uwa, Sakai:2014hsa, Kim:2014dza, Kim:2015jba}.
\begin{figure}
\centering
\includegraphics[width=8cm]{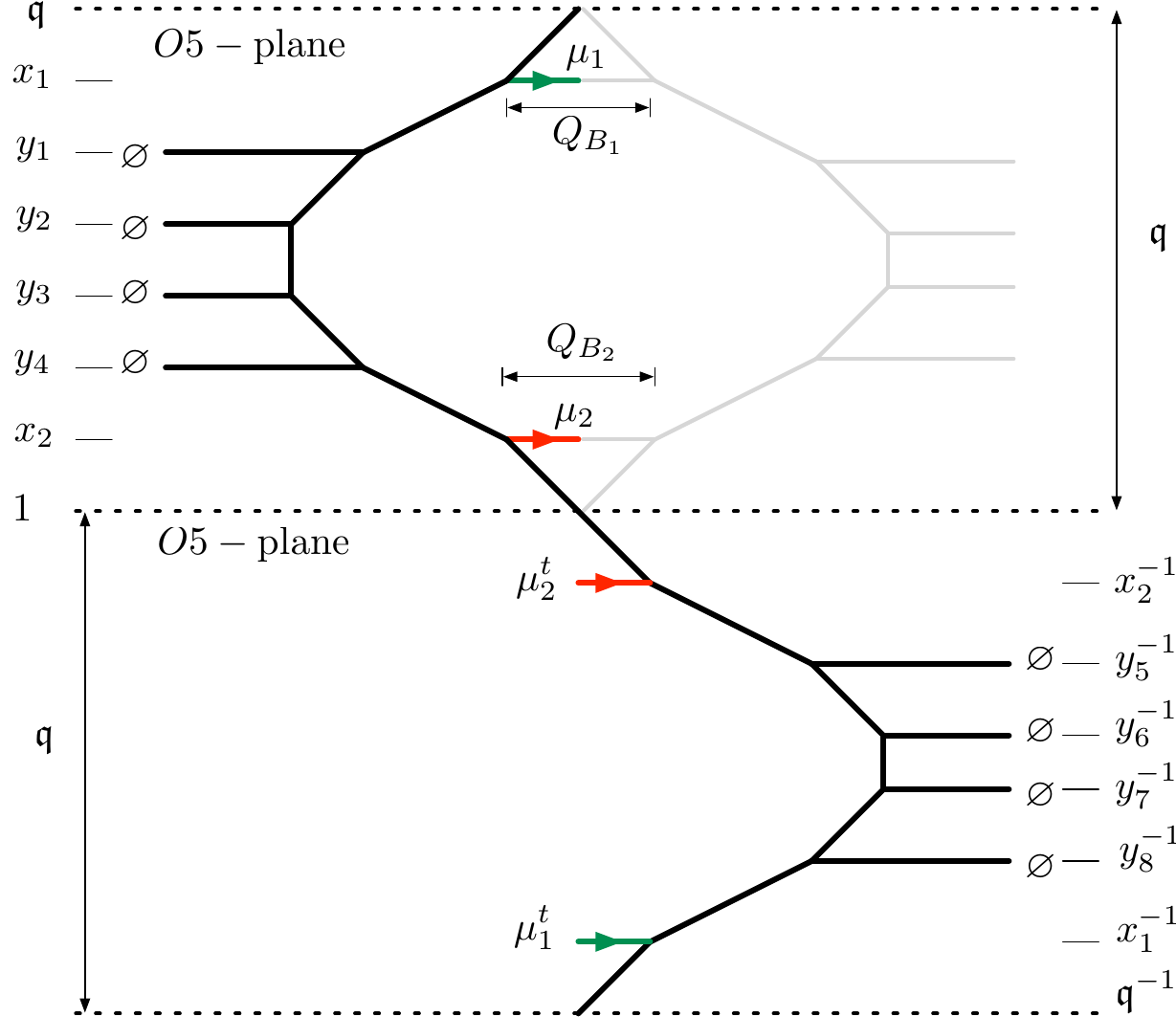}
\caption{A periodic $(p,q)$ brane configuration with two O5-planes for Sp(1) theory with $N_f=8$ flavors, where the periodicity is given as the instanton factor squared $\mathfrak{q}^2$.   The K\"ahler parameters can be easily read off from the positions $x_I,y_i$, for instance, $Q_{B_1} =\mathfrak{q}^2 x_1^{-2}$ and $Q_{B_2} = x_2^2$. Each D5-brane in the middle associated with Young diagrams $\mu_1$ and $\mu_2$ is glued respectively.}
\label{Fig:Estringkahler}
\end{figure}
It is known that the $N_f=8$ case can be realized by two fractional NS5-branes on an O6$^-$-plane in type IIA setup,
whose T-dual picture is 
type IIB $(p,q)$ 5-brane web with two O5-planes as depicted in Fig. \ref{Fig:Estringkahler}.

As two O5-planes are required in Fig. \ref{Fig:Estringkahler}, the covering space for $(p,q)$ 5-brane with O5-planes is periodic as shown in Fig. \ref{Fig:Estringkahler}, where the periodicity is given by the instanton factor squared $\mathfrak{q}^2$.
The periodic strip diagram appearing in Fig. \ref{Fig:Estringkahler} is exactly the one computed in the context of M-string \cite{Haghighat:2013gba,Haghighat:2013tka,Hohenegger:2013ala,Sugimoto:2015nha} 
but with specific tuning of the K\"ahler parameters due to the O5-planes.
For the periodic strip, we replace $R_{\mu \nu}(Q)$ by its infinite product
\begin{align}
	\Theta_{\mu \nu}(Q) 
        \equiv
	\prod_{n=0}^{\infty} R_{\mu\nu}(Q \mathfrak{q}^{2n}) R_{\mu^t \nu^t}(Q^{-1} \mathfrak{q}^{2n+2}),
\end{align}
which yields the topological string partition function for five-dimensional SU(2) gauge theory with $N_f=8$ flavors
\begin{align}\label{eq:Nf8Estring}
Z_8^{\rm O5}&= 
\sum_{\mu_1,\mu_2} \Big( \frac{\mathfrak{q}^2 x_2}{x_1^{3}}
\Big)^{|\mu_1|} \bigg( \frac{\prod_{i=1}^8 y_i}{x_1 x_2^{5}}
\bigg)^{|\mu_2|} 
f_{\mu_1}^{-4}f_{\mu_2}^{-4} 
\\
&\times \prod_{I=1}^{2}\bigg( \prod_{i=1}^{8}
\frac{\Theta_{\mu_I \varnothing}(x_I y_i^{-1})}{\Theta_{\mu_I \varnothing}(x_I y_i)}
\prod_{J=1}^{2}
\frac{\Theta_{\mu_I \mu_J}(x_I x_J)}{\Theta_{\mu_I \mu_J^t}(x_I x_J^{-1})}\bigg),\nonumber
\end{align}
where $x_I, y_i$ are the labels of the location on the $(p,q)$ web diagram with two O5-planes in Fig. \ref{Fig:Estringkahler}. 
Here, we used the analytic continuation corresponding to the flop transition:
\begin{align}
\Theta_{\mu \nu}(Q) \to Q^{|\mu|+|\nu|} f_{\mu} f_{\nu} \Theta_{\mu^t \nu^t}(Q^{-1}).
\end{align}
The parametrization is given by
\begin{align}\label{eq:para}
x_1 = \Lambda\, A, \quad x_2 = \Lambda\, A^{-1}, \quad
y_i = \Lambda\, M_i,
\end{align}
where
$M_i=e^{- \beta m_i}$ are eight mass parameters and 
$ 
\Lambda \equiv \mathfrak{q}^{\frac12} \prod_{i=1}^8 M_i^{-\frac14}
$
\footnote{We note that the parametrization \eqref{eq:para} is the same parametrization discussed in \cite{Hayashi:2016abm}, accounting the parameter map between five-dimensional Sp($N$) gauge theory and SU($N+1$) gauge theory \cite{Gaiotto:2015una}.
}.
We have dropped the factors like $\Theta_{\varnothing \varnothing} (y_i^{\pm 1} y_j^{\pm 1})$,
which do not depend on the Coulomb modulus $A$,
as they are part of the ``extra factor'' \cite{Bergman:2013ala, Hayashi:2013qwa, Bao:2013pwa, Hwang:2014uwa}.
Taking into account the extra factors as well as allowing the flop transitions in the perturbative part 
\footnote{For comparison, we first rewrite \eqref{eq:Nf8Estring} in a plethystic exponential form, whose exponent is given as the expansion in terms of the instanton factor $\mathfrak{q}$. The $\mathfrak{q}^0$ term in this expansion is identified as the perturbative part. 
In order to see the agreement of the perturbative part, we need to change the K\"ahler parameters in some of the terms as $Q \to Q^{-1}$ by hand, which is interpreted as the flop transition. The $A^0$ term in the exponent is called ``extra factor'' and we remove such contributions. 
This is the same technique used e.g. in \cite{Kim:2015jba, Hayashi:2016abm}.},
 we found that our result \eqref{eq:Nf8Estring} is in agreement with the known partition function \cite{Kim:2015jba}. Our proposal hence provides a new, yet simple, expression for $\mathbb{R}^4\times T^2$ partition function of six-dimensional E-string theory.

\section{Conclusions and discussions}\label{sec:conclusion}
In this paper, we proposed a way to implement the topological vertex formulation for ($p,q$) 5-brane configurations with an O5-plane.  The key idea is that, based on a special phase of the brane configuration where 5-branes stuck on an O5-plane meet at a point on the O5-plane, we assign an identical Young diagram to two such 5-branes and introduce the new edge factor \eqref{eq:O5assign} corresponding to such 5-brane configuration.

To test our proposal,
we considered 5-brane webs for the five-dimensional Sp(1) theory with $N_f=0,1$ flavors, 
compared the partition functions computed based on our proposal with the known SU(2) partition functions. For each case, we checked these two partition functions by expanding them in terms of the instanton factor, and found that they do agree up to 10-instanton order.
Another nontrivial check we did is the five-dimensional Sp(1) theory with $N_f=8$ flavors. As it is six-dimensional E-string theory on a circle, the brane configuration consists of two O5-planes and hence naturally shows periodic structure. We found that our partition function for the $N_f=8$ case also agrees with the known E-string elliptic genus partition function. 

It is feasible to apply our method to higher rank Sp($N$) gauge theory which then gives rise to the dual SU($N+1$) theory, and, hence, confirm the duality \cite{Gaiotto:2015una} between two theories in a more manifest way.

Our formalism is also valid even for the S-dual descriptions of $(p,q)$ 5-brane configuration with an O5-plane, which lead to the 5-brane web with an ON-plane \cite{Hanany:1999sj, Hayashi:2015vhy, Zafrir:2015ftn}.
In particular, the web diagram proposed in \cite{Hayashi:2015vhy}
as the ``microscopic'' description of an ON$^0$-plane, 
which is used to construct D-type quiver gauge theories,
 is exactly the S-dual of the case $p=0$ in Fig. \ref{Fig:new1}.
Therefore, it should be straightforward to reproduce the (unrefined) Nekrasov partition function for D-type quiver gauge theories using our proposal
\footnote{
In preparing this paper, we became aware that the partition function for D-type quiver gauge theories is considered in~\cite{Bourgine:2017rik} based on a similar setup.}.

Our proposal may enable one to compute various topological string partition computations where the conventional topological vertex method is not applicable, such as five-dimensional SO($M$) gauge theories with hypermultiplets in vector as well as spinor representations \cite{Hayashi:2018bkd} based on the web diagram proposed in \cite{Zafrir:2015ftn}. 

Finally, it would be useful to extend our method to the refined topological vertex, as the computations in this paper are done based on an unrefined version of topological vertex formulation.


\begin{acknowledgments}
We thank Hirotaka Hayashi and Kimyeong Lee for useful discussions. 
We also thank the authors of \cite{Bourgine:2017rik}
for kindly agreeing to coordinate our submission.  
We would like to thank the
2017 Aspen Winter Conference ``Superconformal Field Theories in $d \ge 4$'' at the Aspen Center for Physics, 
the YITP workshop (YITP-W-17-08) ``Strings and Fields 2017'' at Kyoto University, and ``Tsinghua Summer Workshop in Geometry and Physics 2017" at YMSC Tsinghua University. 
We also would like to thank KIAS for hospitality during visit.
S.S.K. is supported by UESTC Initial Research Grant No.~A03017023801317.
F.Y. is supported in part, by the Israel Science Foundation under Grant No. 352/13.
\end{acknowledgments}


\bibliographystyle{apsrev4-2}
\providecommand{\href}[2]{#2}
\begingroup
\raggedright

\endgroup

\end{document}